\documentclass[conference]{IEEEtran}

\usepackage{graphicx}
\usepackage{float}
\usepackage{tabularx}
\usepackage{cite}
\usepackage{siunitx}
\usepackage{amsmath}
\usepackage[hidelinks]{hyperref}

\begin{document}
\title{Coexistence of Mobile Broadband IMT Systems and UWB Keyless Entry Systems above 6.5 GHz}

\makeatletter
\newcommand{\linebreakand}{
\end{@IEEEauthorhalign}
\hfill\mbox{}\par
\mbox{}\hfill\begin{@IEEEauthorhalign}
}
\makeatother

\author{
\IEEEauthorblockN{Carsten Monka-Ewe\IEEEauthorrefmark{1}, Lukas Berkelmann\IEEEauthorrefmark{2}, Richard Wolf, \\Jan Oliver Oelerich, Thorsten Sch\"onfelder, Marco K\"uhnel, Zhichao Chen, Bert Jannsen}
\IEEEauthorblockA{Volkswagen AG, Wolfsburg, Germany\\
Email:
\IEEEauthorrefmark{1}carsten.monka-ewe@volkswagen.de,
\IEEEauthorrefmark{2}lukas.berkelmann@volkswagen.de}

\linebreakand

\IEEEauthorblockN{Alin Stanescu}
\IEEEauthorblockA{Volkswagen Infotainment GmbH, Bochum, Germany}

\and

\IEEEauthorblockN{Oliver Kushova, Daniel Siekmann}
\IEEEauthorblockA{HELLA GmbH \& Co. KGaA, Lippstadt, Germany}

}

\maketitle

\begin{abstract}
The ever-increasing need for spectrum for mobile broadband systems has led to the recent allocation of spectral resources for International Mobile Telecommunication (IMT) services in the upper mid band (6.425 - 7.125~GHz) at the World Radio Conference (WRC-23) as well as to the creation of an agenda item on identifying future IMT bands in the frequency region 7.125 - 10.5 GHz  at WRC-27. The severity of the impact of these frequency allocations on existing UWB systems, which have been using this part of the spectrum as a sub-secondary user for many years, is still subject to controversial discussions.

This paper contributes a study on the impact of IMT on a real-world vehicular UWB keyless entry system to this discussion. It is shown that both the car's wireless on-board unit and a nearby basestation may drastically affect the system's performance.

\end{abstract}

\IEEEpeerreviewmaketitle

\section{Introduction}
\label{sec:introduction}
Ultra Wideband (UWB) radio systems are a key enabler for a wide variety of applications that require accurate ranging and localization. Especially the release of the IEEE 802.15.4z standard specifying a PHY and MAC layer for UWB communications has recently promoted the widespread use of UWB in consumer products. In the automotive sector, one of the most prominent UWB applications are keyless entry systems. These systems rely on radio communication at rather low frequencies (125 kHz, 433 MHz) while at the same time leveraging UWB ranging to determine the distance between a vehicle and a keyfob to prevent relay station attacks on the system. Figure \ref{fig:uwb_channels} provides an overview of the UWB frequency bands typically used in vehicular access applications. 

From a regulatory perspective, UWB systems are considered sub-secondary users in these bands and consequently have to accept interference caused by other radio services \cite{ECC2022}.

The ever-increasing need for spectrum for mobile broadband systems \cite{GSMA2023} such as 5G and 6G (International Mobile Telecommunications (IMT) standards) as well as for wireless local area networking (Wi-Fi 6E / IEEE~802.11ax) causes a push to allocate spectral resources in the frequency ranges currently used in UWB applications.

In particular, GSMA claims that 2 GHz of mid-band spectrum will be required for IMT by 2030 and thus advocates for the allocation of spectrum in the 6.425 - 7.125 GHz upper mid band and the 7.125 - 10.5 GHz band ("Future IMT") \cite{GSMA2023}.

As a result, at the world radio conference 2023 (WRC-23) a 6 GHz band was established for IMT applications \cite{WRC232023}. In addition, a new agenda item was introduced for WRC-27,\linebreak{} which aims at identifying IMT bands above 7.125 GHz \cite{WRC2023}. 

With the emergence of the extended frequency range for \linebreak{} Wi-Fi (6E / IEEE 802.11ax), numerous studies on the coexistence with UWB systems have been conducted \cite{H.Brunner2022,ECCReport2019}, giving an outlook on the negative impact of other services using the same frequency band. As shown in \cite{H.Brunner2022}, the coexistence with Wi-Fi 6E alone already can cause a significant degradation of both UWB communication and ranging performance. Similarly, a theoretical study carried out for ECC Report 302\cite{ECCReport2019} showed that a significant sensitivity degradation of UWB systems occurs even at relatively large separation distances to Wi-Fi interferers in the same frequency band. However, a Monte-Carlo simulation for an UWB keyless entry system in an urban scenario suggests that the probability of sensitivity degradation is less than \SI{1}{\%} if other factors such as the Wi-Fi system's activity factor and building losses are considered \cite{ECCReport2019}.

Regarding the coexistence of UWB and IMT according to the WRC-23 decision, a similar theoretical study conducted on behalf of FIRA concludes that severe UWB sensitivity degradation is very likely (75 \% and 95 \% of the events for outdoor macro urban and indoor small cells scenario) \cite{Fira2023}. 

The theoretical influence of IMT on UWB systems is thus beyond question. This paper contributes a study of the impact of future 5G/6G IMT systems on a real-world vehicular keyless entry system which transfers the theoretical considerations into a practical application.

\vspace{-0.2cm}
\begin{figure}[H]
	\centering
	\includegraphics[width=0.5\textwidth]{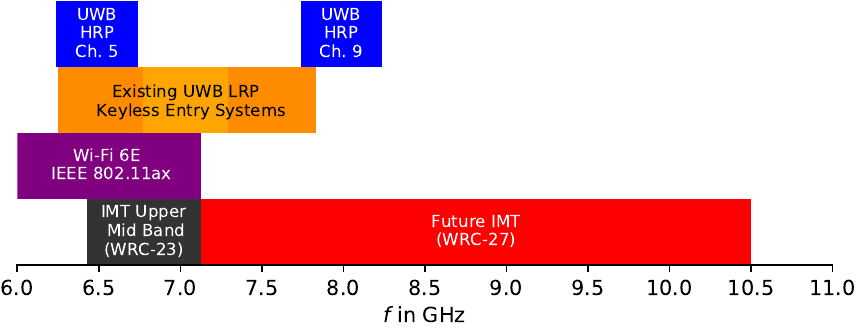}
	\vspace{-0.9cm}
	\caption{UWB channels used in vehicular keyless entry applications. \newline IEEE 802.15.4z HRP channels are shown in blue. Channels used in existing UWB LRP systems are indicated in orange.} 
	\label{fig:uwb_channels}
	\end{figure}

\newpage
\section{Conducted Measurements}
\label{sec:conducted_measurements}
\subsection{Test Setup and Methodology}
In order to study the impact of various IMT interferers on an UWB receiver, two signal generators were used to play back the UWB reference waveform and the IMT interferer waveform. The two generators' signals were combined using a resistive combiner. In contrast to the series-production UWB modules featuring an integrated antenna used in the over-the-air measurements presented in Sections \ref{sec:self-interference-scenario} and \ref{sec:basestation_scenario}, a connectorized UWB module was used as device under test (DUT). Power levels were calibrated at the SMA connector to be attached to the DUT using a spectrum analyzer. The interferer power is given as channel power while the UWB power level is given as peak power as defined in Section 7.4.4. of \linebreak{} ETSI EN 303 883.

The UWB reference waveform was created using a proprietary MATLAB script. Since 6G waveforms are not yet specified and no commercial software is available for signal generation, 5G waveforms created using R\&S WinIQSim2 were used as IMT interferer waveforms. Table \ref{tab:waveforms} summarizes the parameters used for creating the different waveforms.

Successful reception of the UWB reference waveform is reported via the DUT's CAN interface. 

\subsection{Results}
Figure \ref{fig:conducted_measurement} illustrates the impact of various IMT waveforms on the receive sensitivity of the UWB module under test. It can be seen that a sensitivity degradation of $\SI{3}{dB}$ can be observed at interferer power levels as low as $P_\text{Int}=\SI{-75}{dBm}$. This is in good agreement with the $P_\text{Int}=\SI{-78}{dBm}$ reported to cause a \SI{3}{dB} sensitivity degradation in \cite{ECCReport2019,Fira2023}. 

IMT signals with larger bandwidths are found to have a worse impact on the UWB system's receive sensitivity while there is no difference between the impact of IMT uplink and downlink signals.
\vspace{-0.5cm}
\newcolumntype{C}[1]{>{\centering\arraybackslash}p{#1}}
\begin{table}[H]
	\caption{IMT Interferer Waveforms Considered in this Study}
	\vspace{-0.2cm}
		\begin{tabularx}{0.5\textwidth}{|C{0.87cm}|C{1.5cm}|C{1.7cm}|C{1cm}|C{1.8cm}|}
			\hline
			\textbf{Wave}&\multicolumn{4}{|c|}{\textbf{NR FR1 FDD Waveform Parameters }} \\
			\cline{2-5} 
			\textbf{form} & \textbf{\textit{DL / UL$^{\mathrm{a}}$}}& \textbf{\textit{Channel BW }}& \textbf{\textit{SCS}} & \textbf{\textit{RB Allocation}}\\
			\hline
			1 & DL & 100 MHz & 30 kHz& 273@0\\
			2 & UL & 100 MHz & 30 kHz& 270@0\\
			3 & UL & 10 MHz & 30 kHz& 12@6\\
			\hline
			\multicolumn{4}{l}{$^{\mathrm{a}}$UL waveforms use transform precoding.}
		\end{tabularx}
\label{tab:waveforms}
\end{table}

\vspace{-0.6cm}
\begin{figure}[H]
	\centering
	\includegraphics[width=0.47\textwidth]{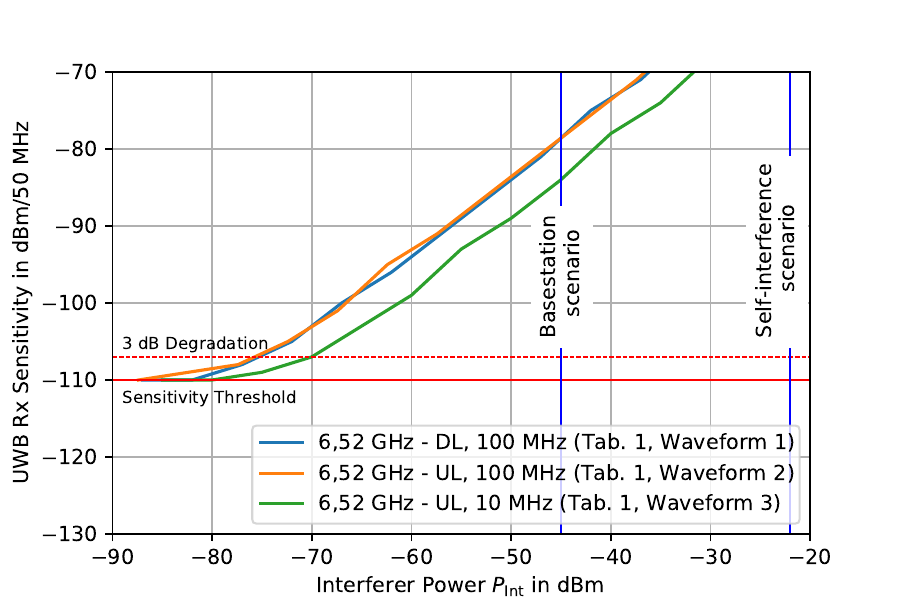}
	\vspace{-0.3cm}
	\caption{Impact of an IMT interferer on the Rx sensitivity of an UWB chipset.}
	\label{fig:conducted_measurement}
\end{figure}

\newpage

\section{Self-interference scenario}
\label{sec:self-interference-scenario}
\subsection{Reasoning}
Today's vehicles are equipped with a wireless on-board unit to provide connectivity to IMT networks in order to implement safety features such as eCall as well as to account for the trend towards deploying over-the-air updates and implementing always-on connectivity features (such as remotely controlling the vehicle's comfort features via a smartphone application). Typically, multiple radiators for IMT connectivity can be found on a modern vehicle, with some radiators being integrated into the roof antenna and some radiators being realized as a separate antenna module hidden in the outer area of the vehicle's body.

At the same time, UWB modules of the vehicle's keyless entry system (featuring an internal radiator) are also often hidden in the same area as is illustrated in Fig.~\ref{fig:self_interference_scenario}.

While the coupling between these two antennas is difficult to predict due to the interaction with the vehicle's body, exemplary measurements indicate that the coupling factor $CF$ between the two antennas may be as high as $CF=\SI{-45}{dB}$.

Taking into account that the wireless on-board unit may transmit with up to $P_\text{Tx}=+\SI{23}{dBm}$ (cf. 3GPP 36.521-1, 3GPP 38.521-1), it immediately becomes obvious that the uplink signal sent out by the vehicle may severely affect the UWB keyless entry system, as it may be as strong as \linebreak{} $P_\text{Int}=P_\text{Tx}+CF=\SI{-22}{dBm}$ (cf. Fig. \ref{fig:conducted_measurement}). The self-interference scenario aims at investigating this effect.

\subsection{Test Setup and Methodology}
The vehicle under test shown in Figure \ref{fig:self_interference_scenario} is equipped with four UWB modules, with one module located in each corner of the vehicle. The keyfob is attached to a partly absorbing body phantom as is illustrated in Figure \ref{fig:self_interference_scenario}. The key was placed on the side of the phantom facing away from the vehicle (backpocket configuration). All UWB modules are connected to the test automation via the vehicle's CAN bus. Ranging can be triggered using a dedicated interface to the test automation.

Since the IMT interferer signal needs to be radiated at frequencies well above the frequency range currently allocated for IMT, a wideband monopole reference antenna designed and manufactured by IMST GmbH, Kamp-Lintfort, was mounted to the vehicle's bumper. 

Waveform 3 from Tab. \ref{tab:waveforms} was used in this scenario.
\vspace{-0.7cm}
\begin{figure}[H]
	\centering
	\includegraphics[width=0.5\textwidth]{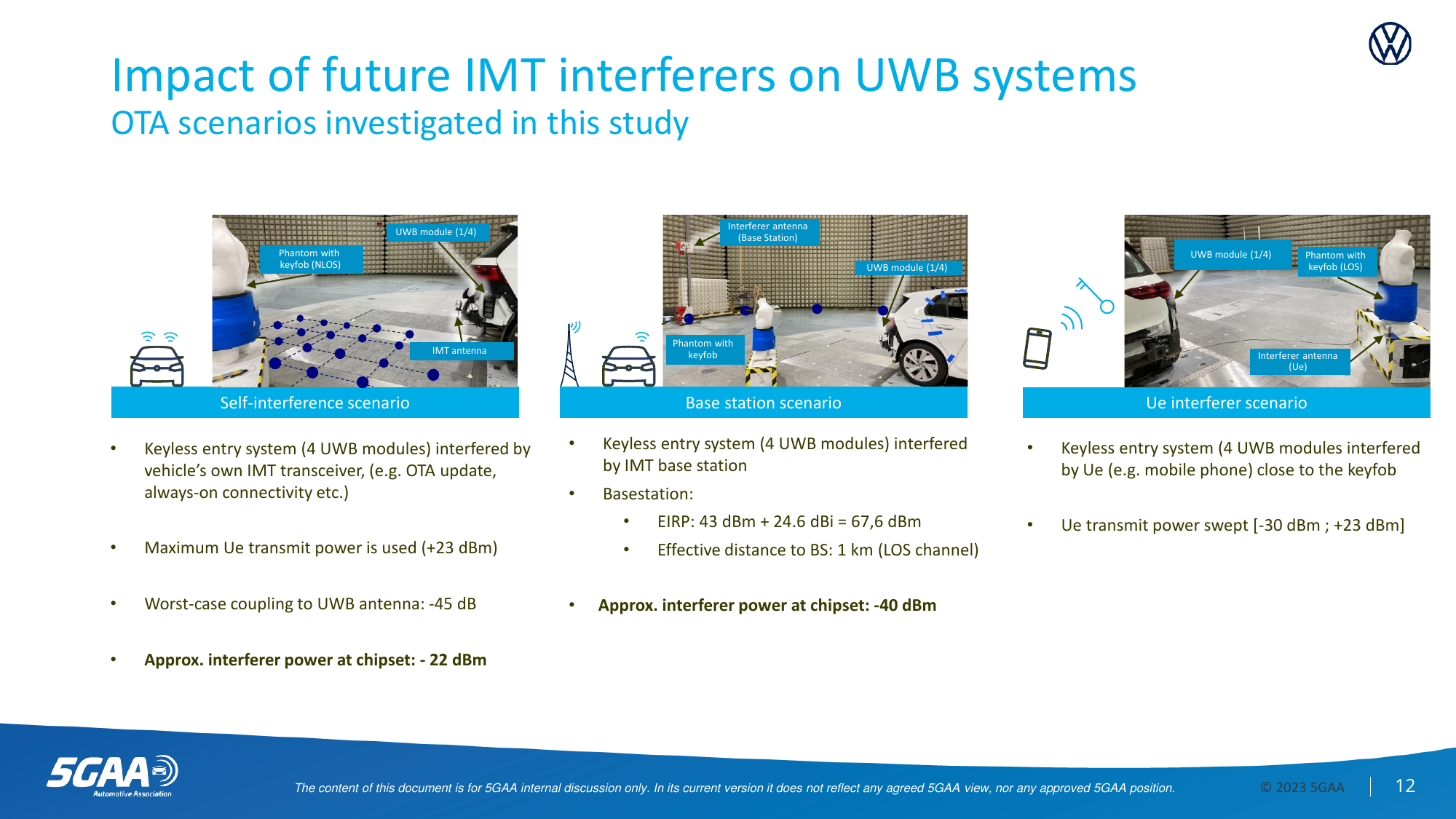}
	\caption{Self-interference scenario. Blue dots indicate measurement positions.}
	\label{fig:self_interference_scenario}
\end{figure}

\newpage
The UWB signal's power level at the UWB modules strongly varies with the position of the keyfob around the car. This is for two reasons: 

Firstly, due to the influence of the vehicle bodywork, the UWB antennas are no longer omnidirectional and have a radiation pattern that varies greatly over the angle. 

Secondly, interference between multiple propagation paths may lead to rapidly fluctuating power levels at the UWB module.

\subsection{Statistical Analysis}
\label{sec:statistical_analysis}

In order to account for this, the UWB system's performance is evaluated over a grid of $M=72$ positions depicted in Figure~\ref{fig:measurement_grid}. At each position, $N=25$ rangings are performed to all four UWB modules simultaneously. If the ranging to at least one UWB module was successful, the ranging attempt is counted as successful. In order to assess the system's performance at a given position $j$, the \textit{Ranging Error Rate} (RER)
\begin{equation}
RER_{j}=\frac{n_j}{N}
\label{eq:rer}
\end{equation}
is introduced as a figure of merit, where $n_j$ denotes the number of failed rangings at position $j$.

In order to quantify the UWB system's performance over the entire grid, we firstly calculate the probability mass function
\begin{equation}
p_{RER}(x_i)=P(RER=x_i)=\frac{m_{RER}}{M}
\end{equation}
where $m_{RER}$ represents the number of observations of discrete values $x_i$ for $RER_j$ over the grid's $M$ positions. In a second step, we calculate the cumulative distribution function (CDF)

\begin{equation}
\label{eq:cdf}
CDF_{RER}(x) 
=
P(RER\leq x)
=
\sum_{x_i\leq x} p_{RER}(x_i).
\end{equation}
Equation (\ref{eq:cdf}) provides the spatial probability for the $RER$ being less than or equal to $x$ over the grid of $M$ positions. Because from a performance point of view it makes more sense to carry out an analysis in terms of a \textit{ranging success rate} $RSR=1-RER$, in the following, we calculate and assess

\begin{equation}
\label{eq:RSR}
P(RSR\geq x) = P(RER\leq x).
\end{equation}
\begin{figure}[H]
	\centering
	\includegraphics[width=0.5\textwidth]{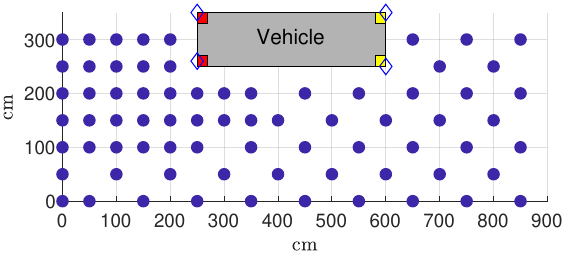}
	\vspace{-0.5cm}
	\caption{Additional representation of the measurement grid used in the self-interference scenario shown in Fig. \ref{fig:self_interference_scenario}. UWB modules are indicated as diamonds. Vehicle width is not shown to scale.}
	\label{fig:measurement_grid}
\end{figure}

\newpage
\section{Basestation scenario}
\label{sec:basestation_scenario}
\subsection{Reasoning}
In \cite{Fira2023}, $EIRP_\text{BS}=\SI{+67.6}{dBm}/\SI{100}{MHz}$ is provided as a reference value for an IMT basestation in a macro urban cell scenario. If we assume the UWB module's internal antenna to have a gain of $G=\SI{+2}{dBi}$ at \SI{6.52}{GHz}, we can estimate the power of the IMT interferer signal at the UWB chipset to be on the order of $P_\text{Int}=\SI{-45}{dBm}$ for a distance of \SI{2000}{m} by means of Friis' formula if a line-of-sight (LOS) channel is assumed.

By comparing this value to the conducted measurements depicted in Figure \ref{sec:conducted_measurements}, it immediately becomes obvious that even at larger distances, a basestation may cause significant impairments of the UWB system's performance. This second scenario aims at investigating this effect in greater detail.

\subsection{Test Setup and Methodology}
As is illustrated in Figure \ref{fig:base_station_scenario}, the IMT downlink interferer signal is radiated by an antenna mounted on an auxiliary mast positioned at $r'=\SI{10}{m}$ away from the vehicle. In this scenario, waveform 1 from Tab. \ref{tab:waveforms} was used.

Calibration of the IMT power level at the vehicle can be performed by equating the radiation density $S_\text{BS}$ created by the basestation at a distance $r$ to the radiation density $S_\text{Int}$ created by the interferer antenna at a distance $r'$, i.e.
\begin{equation}
S_\text{BS}=\frac{EIRP_\text{BS}}{4\pi r^2} = \frac{P_\text{Tx} G}{4 \pi r'^2}=S_\text{Int}
\end{equation}
where $P_\text{Tx}$ denotes the power at the interferer antennas feedpoint and $G$ represents the gain of this antenna. By rearranging this equation as
\begin{equation}
P_\text{Tx}=\frac{r'^2}{r^2}\frac{EIRP_\text{BS}}{G}
\end{equation}
the required power level to be fed into the interferer antenna can be determined for a given distance $r$ to the basestation. The results presented in this paper were obtained for $r=\SI{2000}{m}$.

The position of the keyfob was determined by choosing a grid point which was found to be susceptible to interference in the self-interference scenario. During the entire measurement, the position of the keyfob is not changed. 

\begin{figure}[H]
	\centering
	\includegraphics[width=0.5\textwidth]{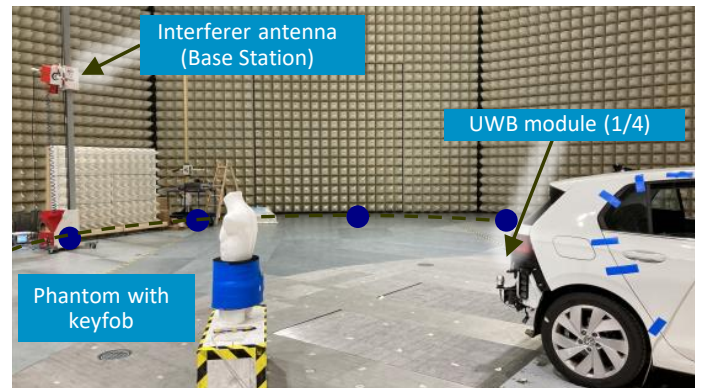}
	\vspace{-0.5cm}
	\caption{Base station interference scenario. Blue dots indicate effective interferer positions.}
	\label{fig:base_station_scenario}
\end{figure}

\newpage

For the same reasons as outlined in Sec. \ref{sec:self-interference-scenario}, the power level of the IMT interferer strongly varies with the position of the interferer antenna. To account for this, the UWB system's performance is evaluated for interferer incidence angles all around the car ($M=36$, 10 degree stepwidth) by rotating the car using the semi-anechoic chamber's turntable.

\subsection{Statistical Analysis}
Statistical analysis of the data is performed in the same way as described in Sec. \ref{sec:statistical_analysis}. In (\ref{eq:rer}), $n_j$ now denotes the number of failed rangings at an interferer incidence angle $j$. The analysis is again performed in terms of $P(RSR\geq x)$ as per (\ref{eq:RSR}), which now represents the spatial probability of the $RSR$ being larger than or equal to $x$ over all interferer incidence angles.

\section{Results and Discussion}
\label{sec:results}
To facilitate the discussion of the UWB keyless entry system's performance, we define the requirement $RSR \geq 0.9$ for a single position as our key performance indicator (KPI).

As can be seen from Figure \ref{fig:results_self_interference}, in the self-interference scenario, this KPI is maintained at approx. $p=\SI{90}{\%}$ of the positions on the grid depicted in Figure \ref{fig:self_interference_scenario} and \ref{fig:measurement_grid} if the interferer is switched off. The fact that less than $p=\SI{100}{\%}$ of the positions maintain the KPI is due to the very tight link budget in the backpocket configuration even with no interferer present. If the interferer is switched on, approx. $p=\SI{25}{\%}$ of the positions maintain the KPI only and consequently, the system's performance has to be considered to be severely affected.

Similar results were obtained for the basestation scenario as is shown in Figure \ref{fig:results_bs_interference}. If the interferer is switched off, $RSR=1$ is maintained at the given keyfob position at all times. Once the interferer is switched on, the KPI is maintained at \linebreak{}$p=\SI{15}{\%}$ of the measured interferer incidence angles only. 

It is important to mention that the interferer waveforms specified in Tab. \ref{tab:waveforms} are FDD waveforms while IMT will rely on TDD signals in the frequency range under consideration. This choice was deliberately made in order to avoid introducing additional statistical effects due to the timing alignment between the IMT and UWB signals to this study.

In light of this, one might argue that less severe degradation would have been observed using TDD waveforms because the UWB system might operate in the IMT downlink's or uplink's transmission gaps. However, because the study shows that both downlink and uplink signals may affect the UWB system, these assumed gaps do not necessarily exist as either of the two signals is present at any point in time.

\section{Conclusion and Future Work}
\label{sec:conclusion}
In this paper, we presented both conducted and over-the-air measurements to investigate the impact of possible future IMT interferers on the performance of a vehicular UWB keyless entry system. Despite the limited scope of this study, it was shown that IMT interferers have a severe impact on the performance of UWB keyless entry systems.

\newpage
\begin{figure}[H]
	\centering
	\includegraphics[width=0.49\textwidth]{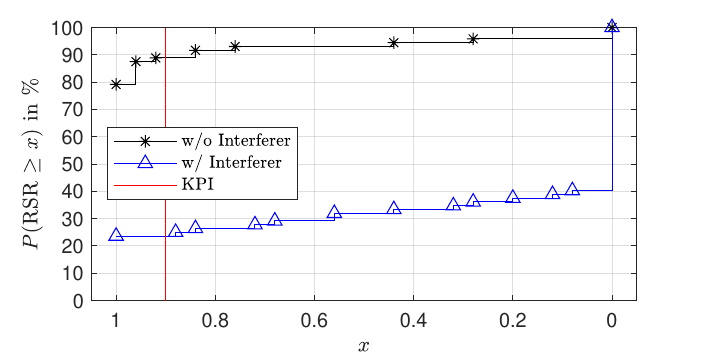}
	\vspace{-0.7cm}
	\caption{Results for the self-interference scenario w/ interferer waveform 3}
	\label{fig:results_self_interference}
\end{figure}
\vspace{-0.65cm} 
\begin{figure}[H]
	\centering
	\includegraphics[width=0.49\textwidth]{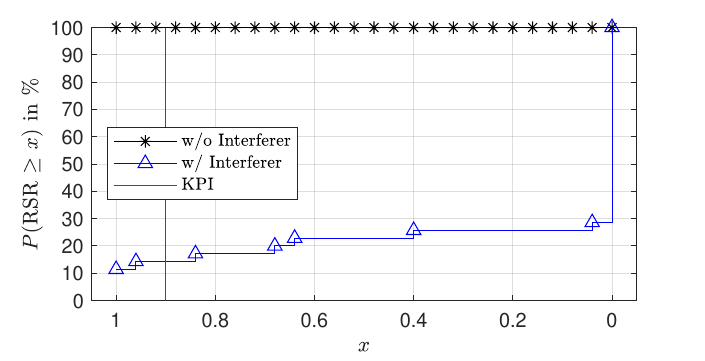}
	\vspace{-0.7cm}	
	\caption{Results for the basestation scenario w/ interferer waveform 1}
	\label{fig:results_bs_interference}
\end{figure}
\vspace{-0.3cm} 

Several other factors such as scheduling, cell loading and beam steering as well as the timing of the UWB rangings have not been considered here, but are likely to have a considerable impact on the two systems' coexistence and thus should be investigated in the context of further measurements. At the same time, there is a need for further theoretical modeling and Monte-Carlo analyses to obtain a more precise assessment of the probability of disturbances in the UWB system. 

\vspace{-0.2cm} 
\bibliographystyle{IEEEtran}
\bibliography{literature} 
\end{document}